\documentclass[12pt]{article}
\usepackage{latexsym}
\usepackage{graphics}
\usepackage{epsfig,amssymb,amsmath,graphicx,amsfonts}
\usepackage[dvips]{color}
\usepackage{hyperref}

\begin{document}

\title{A Note on the Secrecy Capacity of the Multi-antenna Wiretap Channel}
\author{Tie Liu and Shlomo Shamai (Shitz)}
\date{\today}
\maketitle

\begin{abstract}
Recently, the secrecy capacity of the multi-antenna wiretap channel
was characterized by Khisti and Wornell \cite{KW07} using a
Sato-like argument. This note presents an alternative
characterization using a channel enhancement argument. This
characterization relies on an extremal entropy inequality recently
proved in the context of multi-antenna broadcast channels, and is
directly built on the physical intuition regarding to the optimal
transmission strategy in this communication scenario.
\end{abstract}

\section{Introduction}

Consider a multi-antenna wiretap channel with $n_t$ transmit
antennas and $n_r$ and $n_e$ receive antennas at the legitimate
receiver and the eavesdropper, respectively:
\begin{equation}
\begin{array}{lll}
\mathbf{y}_r[m] & = & \mathbf{H}_r\mathbf{x}[m]+\mathbf{w}_r[m] \\
\mathbf{y}_e[m] & = & \mathbf{H}_e\mathbf{x}[m]+\mathbf{w}_e[m]
\end{array}
\label{eq:channel}
\end{equation}
where $\mathbf{H}_r \in \mathbb{R}^{n_r \times n_t}$ and
$\mathbf{H}_e \in \mathbb{R}^{n_e \times n_t}$ are the channel
matrices associated with the legitimate receiver and the
eavesdropper. The channel matrices $\mathbf{H}_r$ and $\mathbf{H}_e$
are assumed to be fixed during the entire transmission and are known
to all three terminals. The additive noise $\mathbf{w}_r[m]$ and
$\mathbf{w}_e[m]$ are white Gaussian vectors with zero mean and are
independent across the time index $m$. The channel input satisfies a
total power constraint
\begin{equation}
\frac{1}{n}\sum_{m=1}^{n}\|\mathbf{x}[m]\|^2 \leq P.
\label{eq:constraint}
\end{equation}
The secrecy capacity is defined as the maximum rate of communication
such that the information can be decoded arbitrarily reliably at the
legitimate receiver but not at the eavesdropper.

For a discrete memoryless wiretap channel $P(Y_r,Y_e|X)$, a
single-letter expression for the secrecy capacity was obtained by
Csisz\'{a}r and K\"{o}rner \cite{CK78} and can be written as
\begin{equation}
C = \max_{P(U,X)} \left[I(U;Y_r)-I(U;Y_e)\right] \label{eq:capacity}
\end{equation}
where $U$ is an auxiliary random variable over a certain alphabet
that satisfies the Markov relation $U - X - (Y_r,Y_e)$. Moreover,
\eqref{eq:capacity} extends to continuous alphabet cases with power
constraint, so the problem of characterizing the secrecy capacity of
the multi-antenna wiretap channel reduces to evaluating
\eqref{eq:capacity} for the specific channel model
\eqref{eq:channel} .

Note that evaluating \eqref{eq:capacity} involves solving a
functional, \emph{nonconvex} optimization problem. Solving
optimization problems of this type usually requires nontrivial
techniques and strong inequalities. Indeed, for the single-antenna
case ($n_t=n_r=n_e=1$), the capacity expression \eqref{eq:capacity}
was successfully evaluated by Leung and Hellman \cite{LH78} using a
result of Wyner \cite{Wyner75} on the degraded wiretap channel and
the celebrated entropy-power inequality \cite[Cha.~16.7]{CT91}.
(Alternatively, it can also be evaluated using a classical result
from estimation theory via a relationship between mutual information
and minimum mean-squared error estimation \cite{GSV}.)
Unfortunately, the same approach does not extend to the
multi-antenna case, as the latter, in its general form, belongs to
the class of \emph{nondegraded} wiretap channels. The problem of
characterizing the secrecy capacity of the multi-antenna wiretap
channel remained open until the recent work of Khisti and Wornell
\cite{KW07}.

In \cite{KW07}, Khisti and Wornell followed an indirect approach to
evaluate the capacity expression \eqref{eq:capacity} for the
multi-antenna wiretap channel. Key to their evaluation is the
following genie-aided upper bound
\begin{eqnarray}
I(U;Y_r)-I(U;Y_e) & \leq & I(U;Y_r,Y_e)-I(U;Y_e) \\
& = & I(X;Y_r,Y_e)-I(X;Y_e)-\left[I(X;Y_r,Y_e|U)-I(X;Y_e|U)\right] \label{eq:markov} \\
& \leq & I(X;Y_r,Y_e)-I(X;Y_e) \label{eq:less noisy} \\
& = & I(X;Y_r|Y_e)
\end{eqnarray}
where \eqref{eq:markov} follows from the Markov chain $U - X -
(Y_r,Y_e)$, and \eqref{eq:less noisy} follows from the trivial
inequality $I(X;Y_r,Y_e|U) \geq I(X;Y_e|U)$. Khisti and Wornell
\cite{KW07} further noticed that the original objective of
optimization $I(U;Y_r)-I(U;Y_e)$ depends on the channel transition
probability $P(Y_r,Y_e|X)$ only through the marginals $P(Y_r|X)$ and
$P(Y_e|X)$, whereas the upper bound $I(X;Y_r|Y_e)$ does depend on
the \emph{joint} conditional $P(Y_r,Y_e|X)$. A good upper bound on
the secrecy capacity is thus contrived as
\begin{equation}
C = \max_{P(U,X)} \left [I(U;Y_r)-I(U;Y_e)\right] \leq
\min_{P(Y_r',Y_e'|X) \in \mathcal{D}} \max_{P(X)} \; I(X;Y_r'|Y_e')
= \max_{P(X)} \min_{P(Y_r',Y_e'|X) \in \mathcal{D}} I(X;Y_r'|Y_e')
\label{eq:upper bound}
\end{equation}
where $\mathcal{D}$ is a set of joint conditionals $P(Y_r',Y_e'|X)$
satisfying
\begin{equation}
P(Y_r'|X)=P(Y_r|X) \quad \mbox{and} \quad P(Y_e'|X)=P(Y_e|X).
\end{equation}
The upper bound $\min_{P(Y_r',Y_e'|X) \in \mathcal{D}} \max_{P(X)}
\; I(X;Y_r'|Y_e')$ has a specific physical meaning: it is the
secrecy capacity of the wiretap channel $P(Y_r',Y_e'|X)$ where the
legitimate user has access to both $Y_r$ and $Y_e$, minimized over
the worst cooperation between the legitimate receiver and the
eavesdropper. In essence, this is very similar to the Sato upper
bound on the sum capacity of a general broadcast channel
\cite{Sato78}. For the multi-antenna wiretap channel, Khisti and
Wornell \cite{KW07} showed that the conditional mutual information
$I(X;Y_r'|Y_e')$ is maximized when the channel input $X$ is
Gaussian. Hence, the upper bounds in \eqref{eq:upper bound} can be
written as a saddle-point matrix optimization problem. By comparing
the value of the optimal \emph{Gaussian} solution for the original
optimization problem $\max_{P(U,X)} [I(U;Y_r)-I(U;Y_e)]$ with the
upper bounds in \eqref{eq:upper bound}, Khisti and Wornell
\cite{KW07} showed that the results are \emph{identical} and thus
established the optimality of both matrix characterizations for the
multi-antenna wiretap channel. Operationally, Khisti and Wornell
\cite{KW07} showed that the original multi-antenna wiretap channel
has the same secrecy capacity as when the legitimate user has access
to both received signals and optimized over the worst cooperation
between the legitimate user and the eavesdropper. (The same approach
was also followed by Shafiee et al. \cite{SNU07} and Oggier and
Hassibi \cite{OH07} to characterize the secrecy capacity of the $2
\times 2 \times 1$ and the general multi-antenna wiretap channel,
respectively.) Considering the disparity between these two physical
scenarios, this is a rather surprising result.

The approach of Khisti and Wornell \cite{KW07} also reminds us of
the degraded same marginals bound for the capacity \emph{region} of
the multi-antenna broadcast channel \cite{VKSJG03, TV03}. There, the
optimality of the Gaussian input is hard to come by, and a precise
characterization of the capacity region had to wait until the
proposal of a drastically different approach by Weingarten et al.
\cite{WSS06}. Motivated by the line of work on the multi-antenna
broadcast channel, in this note we present a different approach to
characterize the secrecy capacity of the multi-antenna wiretap
channel. Our approach is based on an extremal entropy inequality
recently proved in the context of multi-antenna broadcast channels
\cite{LV07, WLSYV07}, and is directly built on the physical
intuition regarding to the optimal transmission strategy in this
communication scenario.

\section{Capacity Characterization via a Channel Enhancement Argument}
\subsection{Capacity characterization}

We consider a canonical version of the channel (vector Gaussian
wiretap channel)
\begin{equation}
\begin{array}{lll}
\mathbf{y}_r[m] & = & \mathbf{x}[m]+\mathbf{w}_r[m] \\
\mathbf{y}_e[m] & = & \mathbf{x}[m]+\mathbf{w}_e[m],
\end{array}
\label{eq:channel2}
\end{equation}
where $\mathbf{x}[m]$ is a real input vector of length $t$, and
$\mathbf{w}_r[m]$ and $\mathbf{w}_e[m]$ are additive Gaussian noise
vectors with zero mean and covariance matrix $\mathbf{K}_r$ and
$\mathbf{K}_e$ respectively and are independent across the time
index $m$. The noise covariance matrices $\mathbf{K}_r$ and
$\mathbf{K}_e$ are assumed to be positive definite. The channel
input satisfies a power-covariance constraint
\begin{equation}
\frac{1}{n}\sum_{m=1}^{n}\mathbf{x}[m]\mathbf{x}^t[m] \preceq
\mathbf{S} \label{eq:constraint2}
\end{equation}
where $\mathbf{S}$ is a positive definite matrix of size $t \times
t$, and ``$\preceq$" represents ``less or equal to" in the positive
semidefinite partial ordering between real symmetric matrices. Note
that \eqref{eq:constraint2} is a rather general constraint that
subsumes many other constraints including the total power constraint
\eqref{eq:constraint}. Following \cite[Sec.~5]{WSS06}, it can be
shown that for any channel gain matrices $\mathbf{H}_r$ and
$\textbf{H}_e$, there exists a sequence of vector Gaussian wiretap
channels \eqref{eq:channel2} whose capacities approach that of the
multi-antenna wiretap channel \eqref{eq:channel}. Without loss of
generality, we shall focus on the vector Gaussian wiretap channel
\eqref{eq:channel2} with power-covariance constraint
\eqref{eq:constraint2} for the rest of the note.

We first present a matrix characterization for the secrecy capacity
of a \emph{degraded} vector Gaussian wiretap channel.

\noindent \textbf{Theorem 1:} If there exists a positive
semidefinite matrix $\mathbf{K}_x^* \preceq \mathbf{S}$ such that
\begin{equation}
\begin{array}{rll}
(\mathbf{K}_x^*+\mathbf{K}_r)^{-1} & = & (\mathbf{K}_x^*+\mathbf{K}_e)^{-1}+\mathbf{M}_2  \\
(\mathbf{S}-\mathbf{K}_x^*)\mathbf{M}_2 & = & 0
\end{array}
\label{eq:kkt1}
\end{equation}
for some positive semidefinite matrix $\mathbf{M}_2$, the secrecy
capacity of a degraded vector Gaussian wiretap channel
\eqref{eq:channel2} with $\mathbf{K}_r \preceq \mathbf{K}_e$ can be
written as
\begin{equation}
C =
\frac{1}{2}\log\det\left(\mathbf{I}+\mathbf{K}_x^*\mathbf{K}_r^{-1}\right)-
\frac{1}{2}\log\det\left(\mathbf{I}+\mathbf{K}_x^*\mathbf{K}_e^{-1}\right).
\label{eq:capacity2}
\end{equation}

Theorem 1 states that if there exists a positive semidefinite matrix
$\mathbf{K}_x^* \preceq \mathbf{S}$ that satisfies \eqref{eq:kkt1},
then $U=\mathbf{X} \sim \mathcal{N}(0,\mathbf{K}_x^*)$ is an optimal
choice for the capacity expression \eqref{eq:capacity} of a degraded
vector Gaussian wiretap channel. Note that this provided a
\emph{sufficient} condition to evaluate optimality for a specific
choice of $(U,\mathbf{X})$. To put in perspective, proving the
optimality of Gaussian $U=\mathbf{X}$ for the degraded vector
Gaussian wiretap channel can be done with relative ease using, for
example, the worst additive noise result of Diggavi and Cover
\cite{DC01}. However, even within the Gaussians, it is not clear how
one could obtain a sufficient condition for the optimal choice of
the covariance matrix, as the matrix optimization problem is (once
again) a \emph{nonconvex} one and the standard Karush-Kuhn-Tucker
(KKT) condition is (a priori) only a \emph{necessary} condition.

\emph{Proof of Theorem 1:} For a degraded wiretap channel
$P(Y_r,Y_e|X)$, Wyner \cite{Wyner75} showed that the secrecy
capacity is given by
\begin{equation}
\max_{P(X)}\left[I(X;Y_r)-I(X;Y_e)\right].
\end{equation}
It thus follows that the secrecy capacity of a degraded vector
Gaussian wiretap channel \eqref{eq:channel2} with $\mathbf{K}_r
\preceq \mathbf{K}_e$ can be written as
\begin{eqnarray}
C & = & \max_{f(\mathbf{X}): \; E[\mathbf{XX}^t] \preceq \mathbf{S}}
\left[I(\mathbf{X};\mathbf{X}+\mathbf{W}_r)-I(\mathbf{X};\mathbf{X}+\mathbf{W}_e)\right]
\\
& = & \max_{f(\mathbf{X}): \; E[\mathbf{XX}^t] \preceq \mathbf{S}}
\left[h(\mathbf{X}+\mathbf{W}_r)-h(\mathbf{X}+\mathbf{W}_e)\right]
-\left(\frac{1}{2}\log\det\mathbf{K}_r-\frac{1}{2}\log\det\mathbf{K}_e\right).
\label{eq:tmp1}
\end{eqnarray}
where $\mathbf{W}_r$ and $\mathbf{W}_e$ are length-$t$ Gaussian
vectors with zero mean and covariance matrix $\mathbf{K}_r$ and
$\mathbf{K}_e$ respectively and are independent of $\mathbf{X}$. As
a special case of Lemma~2 in \cite{WLSYV07}, we have
\begin{equation}
\max_{f(\mathbf{X}): \; E[\mathbf{XX}^t] \preceq \mathbf{S}}
\left[h(\mathbf{X}+\mathbf{W}_r)-h(\mathbf{X}+\mathbf{W}_e)\right]
\leq \frac{1}{2}\log\det\left(\mathbf{K}_x^*+\mathbf{K}_r\right)-
\frac{1}{2}\log\det\left(\mathbf{K}_x^*+\mathbf{K}_e\right).
\label{eq:tmp2}
\end{equation}
(Inequality \eqref{eq:tmp2} was also implicitly used in
\cite[Appendix~C]{LV07}. For completeness, a proof is included in
Appendix~A.) Substituting \eqref{eq:tmp2} into \eqref{eq:tmp1}, we
obtained the desired result \eqref{eq:capacity2}. This completes the
proof. \hfill $\blacksquare$

Next, we use a channel enhancement argument to lift the result of
Theorem 1 to the general vector Gaussian wiretap channel. Channel
enhancement argument was first introduced by Weingarten et al.
\cite{WSS06} to characterize the capacity region of the
multi-antenna broadcast channel. Here, adaptations are made to fit
our purposes. The difference between the channel enhancement
argument here and that of Weingarten et al. \cite{WSS06} will be
explained at the end of Sec.~2.2.

\noindent \textbf{Theorem 2:} The secrecy capacity of a general
vector Gaussian wiretap channel \eqref{eq:channel2} can be written
as
\begin{equation}
C = \max_{0 \preceq \, \mathbf{K}_x \preceq \mathbf{S}}\left[
\frac{1}{2}\log\det\left(\mathbf{I}+\mathbf{K}_x\mathbf{K}_r^{-1}\right)-
\frac{1}{2}\log\det\left(\mathbf{I}+\mathbf{K}_x\mathbf{K}_e^{-1}\right)\right]
\label{eq:capacity3}
\end{equation}
where an optimal $\mathbf{K}_x$ (denoted here as $\mathbf{K}_x^*$)
must satisfy
\begin{equation}
\begin{array}{rll}
(\mathbf{K}_x^*+\mathbf{K}_r)^{-1} + \mathbf{M}_1 & = &
(\mathbf{K}_x^*+\mathbf{K}_e)^{-1}+\mathbf{M}_2
\\ \mathbf{K}_x^*\mathbf{M}_1 & = & 0 \\
(\mathbf{S}-\mathbf{K}_x^*)\mathbf{M}_2 & = & 0
\end{array}
\label{eq:kkt2}
\end{equation}
for some positive semidefinite matrices $\mathbf{M}_1$ and
$\mathbf{M}_2$.

Note that unlike Theorem 1, the characterization \eqref{eq:kkt2} for
the optimal covariance matrix $\mathbf{K}_x$ is based on the
standard KKT condition and hence is only a necessary condition.

\emph{Proof of Theorem 2:} Let $\mathbf{K}_x^*$ be an optimal
solution to the optimization problem in \eqref{eq:capacity3}. By the
KKT condition, $\mathbf{K}_x^*$ must satisfy the equations in
\eqref{eq:kkt2}. Recall the single-letter capacity expression
\eqref{eq:capacity} and let $U=\mathbf{X} \sim
\mathcal{N}(0,\mathbf{K}_x^*)$. The secrecy capacity of a general
vector Gaussian wiretap channel \eqref{eq:channel2} can be bounded
from below as
\begin{equation}
C \geq
\frac{1}{2}\log\det\left(\mathbf{I}+\mathbf{K}_x^*\mathbf{K}_r^{-1}\right)-
\frac{1}{2}\log\det\left(\mathbf{I}+\mathbf{K}_x^*\mathbf{K}_e^{-1}\right).
\label{eq:tmp3}
\end{equation}
To prove the reverse inequality, consider a new vector Gaussian
wiretap channel with legitimate receiver and eavesdropper noise
covariance matrix being $\tilde{\mathbf{K}}_r$ and $\mathbf{K}_e$
respectively, where $\tilde{\mathbf{K}}_r$ is defined through the
equation
\begin{equation}
(\mathbf{K}_x^*+\tilde{\mathbf{K}}_r)^{-1} =
(\mathbf{K}_x^*+\mathbf{K}_r)^{-1} + \mathbf{M}_1. \label{eq:kkt3}
\end{equation}
Following Lemmas 10 and 11 of \cite{WSS06}, $\tilde{\mathbf{K}}_r$
has the following important properties:
\begin{enumerate}
\item $0 \preceq \tilde{\mathbf{K}}_r \preceq \{\mathbf{K}_r,\mathbf{K}_e\}$;
\item $\det(\mathbf{I}+\mathbf{K}_x^*\tilde{\mathbf{K}}_r^{-1})=
\det(\mathbf{I}+\mathbf{K}_x^*\mathbf{K}_r^{-1})$.
\end{enumerate}
By virtue of $\tilde{\mathbf{K}}_r \preceq \mathbf{K}_e$, the new
vector Gaussian wiretap channel is a degraded one. Furthermore, by
the first and third equation in \eqref{eq:kkt2} and \eqref{eq:kkt3}
we have
\begin{equation}
\begin{array}{rll}
(\mathbf{K}_x^*+\tilde{\mathbf{K}}_r)^{-1} & = & (\mathbf{K}_x^*+\mathbf{K}_e)^{-1}+\mathbf{M}_2 \\
(\mathbf{S}-\mathbf{K}_x^*)\mathbf{M}_2 & = & 0.
\end{array}
\end{equation}
It thus follows from Theorem 1 that the secrecy capacity of this new
channel is equal to
\begin{eqnarray}
\tilde{C} & = &
\frac{1}{2}\log\det\left(\mathbf{I}+\mathbf{K}_x^*\tilde{\mathbf{K}}_r^{-1}\right)-
\frac{1}{2}\log\det\left(\mathbf{I}+\mathbf{K}_x^*\mathbf{K}_e^{-1}\right)
\\
& = &
\frac{1}{2}\log\det\left(\mathbf{I}+\mathbf{K}_x^*\mathbf{K}_r^{-1}\right)-
\frac{1}{2}\log\det\left(\mathbf{I}+\mathbf{K}_x^*\mathbf{K}_e^{-1}\right)
\end{eqnarray}
where the last equality is due to the second property of
$\tilde{\mathbf{K}}_r$. Note from the first property of
$\tilde{\mathbf{K}}_r$ that $\tilde{\mathbf{K}}_r \preceq
\mathbf{K}_r$. Reducing the noise covariance matrix for the
legitimate receiver can only increase the secrecy capacity, so we
have
\begin{equation}
C \leq \tilde{C} =
\frac{1}{2}\log\det\left(\mathbf{I}+\mathbf{K}_x^*\mathbf{K}_r^{-1}\right)-
\frac{1}{2}\log\det\left(\mathbf{I}+\mathbf{K}_x^*\mathbf{K}_e^{-1}\right)
\label{eq:tmp4}
\end{equation}
which is the desired reverse inequality. Putting together
\eqref{eq:tmp3} and \eqref{eq:tmp4} completes the proof of the
theorem. \hfill $\blacksquare$

\subsection{Physical intuition}

Our approach of characterizing the secrecy capacity of the vector
Gaussian wiretap channel hinges on the existence of an enhanced
channel, which needs to satisfy:
\begin{enumerate}
\item it is degraded, so the secrecy capacity can be readily
characterized;
\item it has the same secrecy capacity as the original wiretap channel.
\end{enumerate}
A priori, it is not clear whether such an enhanced channel would
always exist, letting alone to actually construct one.

Our intuition regarding to the existence of the enhanced channel was
mainly from the parallel Gaussian wiretap channel, which is a
special case of the vector Gaussian wiretap channel
\eqref{eq:channel2} with \emph{diagonal} noise covariance matrices
$\mathbf{K}_r$ and $\mathbf{K}_e$. In this case, it is shown in
\cite{LVS06} that the optimal transmission strategy is to transmit
only to the subchannels for which the received signal by the
legitimate receiver is stronger than that by the eavesdropper.
Therefore, an enhanced channel can be constructed by reducing the
noise variance for the legitimate receiver in each of those
subchannels to the noise variance level of the eavesdropper.
Clearly, the enhanced channel thus constructed is a degraded
parallel Gaussian broadcast channel. Furthermore, the secrecy
capacity of the enhanced channel is the same as the original
channel, as the noise variances for the legitimate receiver did not
change at all for any of the ``active" subchannels. Therefore, at
least for the special case of the parallel Gaussian wiretap channel,
an enhanced channel does always exist.

Carrying over to the general vector Gaussian wiretap channel, no
information should be transmitted along any direction where the
eavesdropper observes a stronger signal than the legitimate
receiver. The effective channel for the eavesdropper is thus a
degraded version of the effective channel for the legitimate
receiver. (This observation was also made by Khisti and Wornell
\cite{KW07}.) This is the basis underlying the existence of the
enhanced channel for a general vector Gaussian wiretap channel.

Note that in characterizing the capacity region of the vector
Gaussian broadcast channel (a canonical model for the multi-antenna
broadcast channel), Weingarten et al. \cite{WSS06} enhanced each and
every channel (by reducing the noise covariance matrices) from the
transmitter to the receivers. In our argument, however, we only
enhanced the channel for the legitimate receiver. (The channel for
the eavesdropper did not change at all). This is due to the fact
that in both arguments, the enhancement, \emph{a priori}, must
increase the capacity (secrecy or regular) of the channel.
(Otherwise, both arguments will break down.) Whereas reducing the
noise covariances will benefit all the receivers and hence improve
the capacity of the vector Gaussian broadcast channel, reducing the
noise covariance matrix of the eavesdropper may compromise the
security of the transmission scheme and hence lower the secrecy
capacity of the vector Gaussian wiretap channel. This is the key
difference between the channel enhancement argument here and that of
Weingarten et al. \cite{WSS06} for the vector Gaussian broadcast
channel.

\begin{appendix}
\section{Proof of Inequality \eqref{eq:tmp2}}
To prove inequality \eqref{eq:tmp2}, it is equivalent to show that
$\mathbf{X}_G^* \sim \mathcal{N}(0,\mathbf{K}_x^*)$ is an optimal
solution to the optimization problem
\begin{equation*}
\max_{f(\mathbf{X}): \; E[\mathbf{XX}^t] \preceq \mathbf{S}}
\left[h(\mathbf{X}+\mathbf{W}_r)-h(\mathbf{X}+\mathbf{W}_e)\right]
\end{equation*}
which would handle the Gaussianity and the covariance matrix issues
in one shot. For that purpose, we shall prove that $g(\mathbf{X})
\leq g(\mathbf{X}_G^*)$ where
\begin{equation}
g(\mathbf{X}) :=
h(\mathbf{X}+\mathbf{W}_r)-h(\mathbf{X}+\mathbf{W}_e)
\end{equation}
for any $\mathbf{X}$ such that $E[\mathbf{XX}^t] \preceq
\mathbf{S}$.

For any $\mathbf{X}$ such that $E[\mathbf{XX}^t] \preceq \mathbf{S}$
and any $\lambda \in [0,1]$, let
\begin{equation}
\mathbf{X}_\lambda := \sqrt{1-\lambda} \mathbf{X} + \sqrt{\lambda}
\mathbf{X}_G^*
\end{equation}
where we assume that $\mathbf{X}$ and $\mathbf{X}_G^*$ are
independent. By the de-Bruijn identity \cite[Cha.~16.6]{CT91},
\begin{equation}
\frac{d g(\mathbf{X}_\lambda)}{d\lambda} =
\frac{1}{2(1-\lambda)}\mathrm{Tr}\left((\mathbf{K}_x^*+\mathbf{K}_r)\mathbf{J}(\mathbf{X}_\lambda+\mathbf{W}_r)
-(\mathbf{K}_x^*+\mathbf{K}_e)\mathbf{J}(\mathbf{X}_\lambda+\mathbf{W}_e)\right)
\label{eq:tmp5}
\end{equation}
where $\mathbf{J}(\mathbf{X})$ denotes the Fisher information matrix
of $\mathbf{X}$. Recalling the vector Fisher information inequality
\cite[Lemma~1]{WLSYV07}
\begin{equation}
\mathbf{J}(\mathbf{X}_1 + \mathbf{X}_2) \preceq  \mathbf{A}
\mathbf{J}(\mathbf{X}_1)\mathbf{A}^t + (\mathbf{I}-\mathbf{A})
\mathbf{J}(\mathbf{X}_2)(\mathbf{I}-\mathbf{A})^t
\end{equation}
for two independent random vectors $\mathbf{X}_1$ and $\mathbf{X}_2$
and letting
\begin{equation}
\mathbf{A} =
(\mathbf{K}_x^*+\mathbf{K}_e)^{-1}(\mathbf{K}_x^*+\mathbf{K}_r),
\end{equation}
we have
\begin{eqnarray}
\hspace{-18pt} \mathbf{J}(\mathbf{X}_\lambda+\mathbf{W}_r) & \preceq
&
\mathbf{A}^{-1}(\mathbf{J}(\mathbf{X}_\lambda+\mathbf{W}_e)-(\mathbf{I}-\mathbf{A})
\mathbf{J}(\mathbf{W})(\mathbf{I}-\mathbf{A})^t)\mathbf{A}^{-t}
\nonumber
\\
& = &(\mathbf{K}_x^*+\mathbf{K}_r)^{-1}
((\mathbf{K}_x^*+\mathbf{K}_e)\mathbf{J}(\mathbf{X}_\lambda+\mathbf{W}_e)(\mathbf{K}_x^*+\mathbf{K}_e)-
(\mathbf{K}_e-\mathbf{K}_r))(\mathbf{K}_x^*+\mathbf{K}_r)^{-1}
\label{eq:tmp6}
\end{eqnarray}
where $\mathbf{W}$ is $\mathcal{N}(0,\mathbf{K}_e-\mathbf{K}_r)$ and
is independent of $(\mathbf{W}_r,\mathbf{X},\mathbf{X}_G^*)$.
Substituting \eqref{eq:tmp6} into \eqref{eq:tmp5}, we have
\begin{eqnarray}
\frac{d g(\mathbf{X}_\lambda)}{d\lambda} & \geq &
\frac{1}{2(1-\lambda)}\mathrm{Tr}\left((\mathbf{K}_x^*+\mathbf{K}_r)(
\mathbf{J}(\mathbf{X}_\lambda+\mathbf{W}_r)(\mathbf{K}_x^*+\mathbf{K}_r)-\mathbf{I})
((\mathbf{K}_x^*+\mathbf{K}_r)^{-1}-(\mathbf{K}_x^*+\mathbf{K}_e)^{-1})\right)
\nonumber \\
& = &
\frac{1}{2(1-\lambda)}\mathrm{Tr}\left((\mathbf{K}_x^*+\mathbf{K}_r)(
\mathbf{J}(\mathbf{X}_\lambda+\mathbf{W}_r)(\mathbf{K}_x^*+\mathbf{K}_r)-\mathbf{I})\mathbf{M}_2\right)
\label{eq:tmp7} \\
& \geq &
\frac{1}{2(1-\lambda)}\mathrm{Tr}\left((\mathbf{K}_x^*+\mathbf{K}_r)(
(\mathbf{S}+\mathbf{K}_r)^{-1}(\mathbf{K}_x^*+\mathbf{K}_r)-\mathbf{I})\mathbf{M}_2\right)
\label{eq:tmp8} \\
& = &
\frac{1}{2(1-\lambda)}\mathrm{Tr}\left((\mathbf{K}_x^*+\mathbf{K}_r)
(\mathbf{S}+\mathbf{K}_r)^{-1}(\mathbf{K}_x^*-\mathbf{S})\mathbf{M}_2\right)
\nonumber
\\ & = & 0 \label{eq:tmp9}
\end{eqnarray}
where equalities \eqref{eq:tmp7} and \eqref{eq:tmp9} are due to the
equations in \eqref{eq:kkt1}, and inequality \eqref{eq:tmp8} is due
to the well-known Cram\'{e}r-Rao inequality
\begin{equation}
\mathbf{J}(\mathbf{X}) \succeq \mathrm{Cov}^{-1}(\mathbf{X})
\end{equation}
and the fact that $\mathrm{Cov}(\mathbf{X}) \preceq E[\mathbf{XX}^t]
\preceq \mathbf{S}$. That is, $g(\mathbf{X}_\lambda)$ is a
monotonically nondecreasing function of $\lambda$ in $[0,1]$. We
thus have
\begin{equation}
g(\mathbf{X}) = g(\mathbf{X}_0) \leq g(\mathbf{X}_1) =
g(\mathbf{X}_G^*).
\end{equation}
This completes the proof of inequality \eqref{eq:tmp2}.
\end{appendix}

\end{document}